\documentclass[pdftex,12pt]{article}
\usepackage{times}
\usepackage{geometry}
\usepackage[utf8]{inputenc}
\usepackage{enumitem,amssymb}
\usepackage{ragged2e}
\usepackage[numbers,sort&compress]{natbib}
\usepackage{graphicx}
\usepackage{anyfontsize}
\usepackage{wrapfig}
\newlist{thematic}{itemize}{8}
\setlist[thematic]{label=$\square$}
\usepackage{pifont}
\newcommand{\cmark}{\ding{51}}%
\newcommand{\done}{\rlap{$\square$}{\raisebox{2pt}{\large\hspace{1pt}\cmark}}%
\hspace{-2.5pt}}

\setlist[itemize]{itemsep=2pt,wide=10pt,leftmargin=\dimexpr\labelwidth + 3\labelsep\relax,topsep=5pt}

\usepackage{titlesec}

\titleformat*{\section}{\large\bfseries}

\begin{document}
\thispagestyle{empty}
{\raggedright
\huge
Astro2020 Science White Paper \linebreak

Hunting for ancient brown dwarfs: the developing field of brown dwarfs in globular clusters \linebreak
\normalsize

\noindent \textbf{Thematic Areas:} \hspace*{60pt} $\square$ Planetary Systems \hspace*{10pt} $\square$ Star and Planet Formation \hspace*{20pt}\linebreak
$\square$ Formation and Evolution of Compact Objects \hspace*{31pt} $\square$ Cosmology and Fundamental Physics \linebreak
  $\done$  Stars and Stellar Evolution \hspace*{1pt} $\done$ Resolved Stellar Populations and their Environments \hspace*{40pt} \linebreak
  $\square$    Galaxy Evolution   \hspace*{45pt} $\square$             Multi-Messenger Astronomy and Astrophysics \hspace*{65pt} \linebreak
  
\textbf{Principal Author:}

Name: Ilaria Caiazzo
 \linebreak						
Institution: University of British Columbia  
 \linebreak
Email: ilariacaiazzo@phas.ubc.ca
 \linebreak
Phone:  +16047102699
 \linebreak
 
\textbf{Co-authors:}
  Adam Burgasser (UC San Diego),
  Jon M. Rees (UC San Diego),
  France Allard (ENS-Lyon),
  Andrea Dieball (Bonn),
  Jeremy Heyl (UBC),
  Harvey Richer (UBC),
  Isabelle Baraffe (Exeter),
  Christian Knigge (Southampton).
}
\\
\\
\textbf{Abstract:} The detection of brown dwarfs in globular star clusters will allow us to break the degeneracies in age, mass and composition that affect our current models, and therefore to constrain the physics of their atmospheres and interiors. Furthermore, detecting brown dwarfs will help us constraining the properties of the clusters themselves, as they carry information about the clusters' age and dynamics. Great advancements in this field are to be expected in the next ten years, thanks to the extraordinary sensitivity in the infrared of upcoming telescopes like \textit{JWST} and the \textit{ELT}s.

\pagebreak
\pagenumbering{arabic}
\section{Introduction}

Brown dwarfs are among the latest additions to our collection of sky's objects. Even though their existence was first proposed back in the 1960s \cite{1963ApJ...137.1121K,1963PThPh..30..460H}, the first brown dwarfs were not discovered until the 1990s \cite{1995Natur.378..463N,1995Natur.377..129R,1996ApJ...458..600B}. This delay is not due to brown dwarfs being rare; at least 15\% of ``stellar'' objects in the Solar Neighborhood are brown dwarfs \cite{2012ApJ...753..156K}.
The delay is due to brown dwarfs' extreme faintness: unlike stars, brown dwarfs are not massive enough to sustain core hydrogen burning. As a result, a brown dwarf's luminosity is governed by the left-over heat from their formation. As time passes, brown dwarfs cool and dim, and their spectral energy distributions increasingly shift to near-infrared wavelengths. Their discovery required a combination of better search techniques, spectroscopic followup, and observational sensitivity at infrared wavelengths.

Astronomy is currently at a similar turning point: the next generation of space telescopes, such as \textit{JWST}, and ground-based extremely large telescopes, such as \textit{TMT}, \textit{GMT} and \textit{ELT}, will achieve unprecedented sensitivity in the infrared, where brown dwarfs' spectra peak. This will enable the detection and study of brown dwarfs that are colder, older, and further away, including the oldest brown dwarfs in the Galaxy: those found in globular star clusters.  

Members of a globular cluster share to a first degree the same age, chemical composition and distance from the Sun, properties generally determined from observations of stars on the red giant branch, main sequence turn-off or white dwarfs. For the first time, we will have large samples of brown dwarfs for which these fundamental properties are known to high accuracy, allowing us to break many of the observational degeneracies that arise from their cooling nature. For instance, while brown dwarfs in the vicinity of the Sun possess a range of masses and ages, and hence broad diversity in temperatures, luminosities and colors; those in globular clusters lie along a single age- and metallicity-dependent cooling sequence tracing mass. This will enable robust determination of these fundamental parameters and analysis of broader properties such as the substellar mass function, in a manner not currently possible in the field (due to degeneracies) or nearby open clusters (due to small number statistics).

Brown dwarfs occupy a peculiar position as bridges between stars and planets, and understanding their physics has important implications in other fields: from star and planet formation and evolution, to dense-matter physics and galaxy evolution. Furthermore, if properly modeled, brown dwarfs in globular clusters could provide a new method for estimating the age of the clusters themselves. As brown dwarfs age, they become cooler and fainter. As a result, a gap between the end of the hydrogen-burning main sequence and the most massive and hence brightest brown dwarfs is predicted to form \cite{2004ApJS..155..191B,2005ApJ...625..385A,2017arXiv170200091C}. The older the cluster, the more time its brown dwarfs have had to cool, and the wider the gap becomes. Metallicity effects further play a role in the gap location and breadth. 
This previously untapped feature of cluster color-magnitude diagrams offers us a new tool for determining the age of clusters of all kinds.  

Because brown dwarfs in globular clusters have been cooling since almost the beginning of time, the gap is particularly enhanced in these systems. It also means that brown dwarfs in globular clusters are exceedingly faint, currently below the detection limit of the best telescopes. Deep \textit{HST} observations in globular clusters already reach the end of the main sequence \cite{1538-3881-143-1-11,2013ApJ...778..104R}, and while some candidate brown dwarfs have been detected in M4, they remain too faint for follow-up spectroscopic confirmation and characterization \cite{2016ApJ...817...48D}. 

The ability to study brown dwarfs in globular clusters over the next decade will usher in a new way of exploring the astrophysical properties of these unique objects, while in turn providing new opportunities for studying the star formation history of clusters and the Milky Way Galaxy at large.

\section{Globular clusters: a laboratory for brown dwarfs}

Globular clusters are the oldest and most metal-poor stellar aggregates in our Galaxy, with ages typically older than 10 Gyr. As such, they can be seen as ``fossils'' from the violent epoch of Galactic formation, and thus contain information about the earliest stages of star formation in the Milky Way. Stars in a globular cluster belong to a highly uniform population in age and chemical composition, despite subtle differences in composition observed in many systems \cite{2004ARA&A..42..385G,2018ARA&A..56...83B}. Each globular cluster represents an unique snapshot in the evolution of hundreds of thousands of stars, differentiated mainly by their mass. Depending on their mass, stars evolve at different rates, which means that in a single globular cluster we can find stars in virtually every evolutionary stage, from the main sequence to the end point of stellar evolution like white dwarfs and neutron stars. For this reason, globular clusters have served as ideal laboratories for constraining the physics behind stellar evolution.
With brown dwarfs, we can study an entirely distinct evolutionary path.

Physical interpretation of brown dwarfs in the field of the Galaxy are currently limited by uncertainties in age, mass and metallicity. A brown dwarf's observational properties --- temperature, luminosity, spectral type, and detailed spectral and photometric characteristics --- are influenced in complex ways by its age, mass, temperature, composition, cloud content, atmospheric dynamics, etc \cite{1997ARA&A..35..137A,2001ApJ...556..357A}. It is extremely hard to break these degeneracies for individual brown dwarfs, and hence determining these physical properties for individual brown dwarfs remains a major goal for the field. In globular clusters, distance, metallicity and age are already known, leaving properties such as mass to be inferred from the position of an object on the color-magnitude diagram. Being able to simultaneously measure the age, mass and composition for individual brown dwarfs will be a tremendous boon to the field, allowing detailed comparison of evolutionary and structural models to observations, assessment of metallicity-dependent star formation and dynamical scattering processes, and benchmarks for low-temperature, metal-poor atmosphere models. 


Measuring the spectra of brown dwarfs in globular clusters provides a particularly important opportunity to study atmospheric processes common to both these objects and giant exoplanets. At the temperatures that characterize L- and T-type brown dwarfs, molecules dominate gas opacity, while condensate species (including mineral, salt and ice grains) can both redistribute flux toward longer wavelengths and modify gas abundances through rainout \cite{2001ApJ...556..357A}. Gas and grain chemistry and atmospheric dynamics all play essential roles in shaping brown dwarf spectra. The subsolar metallicities and non-solar elemental abundances typical of globular cluster stars means that their brown dwarf spectra are distinct from field counterparts. It has already been seen that L-type halo subdwarfs exhibit distinct chemical peculiarities (enhanced metal hydride abundances) and spectral energy distributions (suppressed K-band flux due to collision-induced H$_2$ absorption) that radically influence their spectra (Figure 1; \citealt{2003ApJ...592.1186B,2006AJ....132.2372G,2017MNRAS.464.3040Z}). Some surprising trends, such as enhanced TiO absorption, can be attributed to changes in cloud formation \cite{2007ApJ...657..494B}. These variations are qualitatively understood, but the lack of independent metallicity or mass determinations for halo subdwarfs has prevented quantitative assessment of spectral trends or robust tests of models \cite{2009ApJ...697..148B}. With globular cluster brown dwarfs, we will be able to properly assess the fidelity of models to reproduce metallicity dependencies in gas and cloud chemistry, while potentially improving determination of cluster metallicity measurements and the segregation of distinct evolutionary tracks. 

\begin{figure}
  \centering
    \includegraphics[width=0.45\textwidth,clip,trim=0.0in 0 0.0in 0]{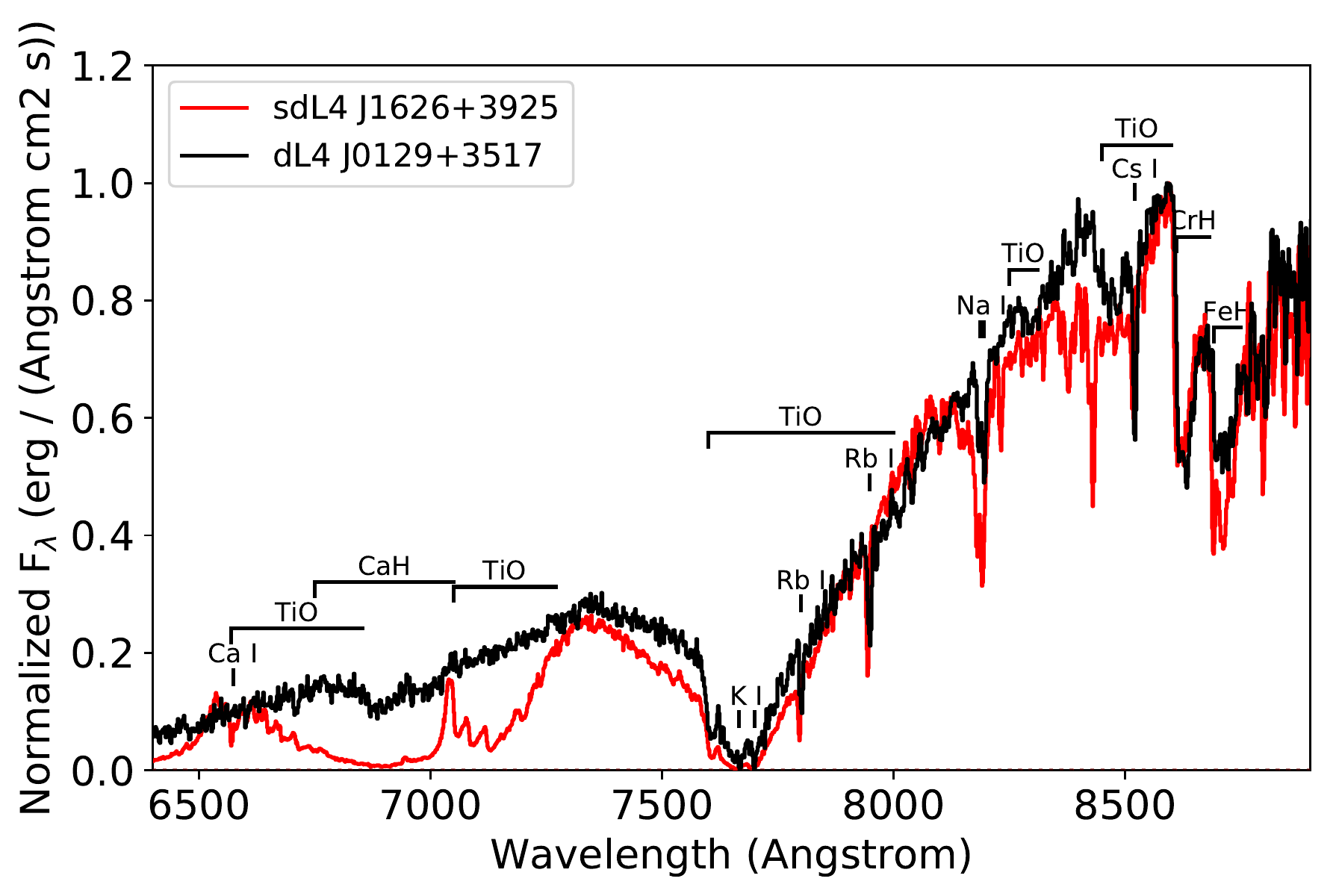}
    \includegraphics[width=0.45\textwidth,clip,trim=0.0in 0 0.0in 0]{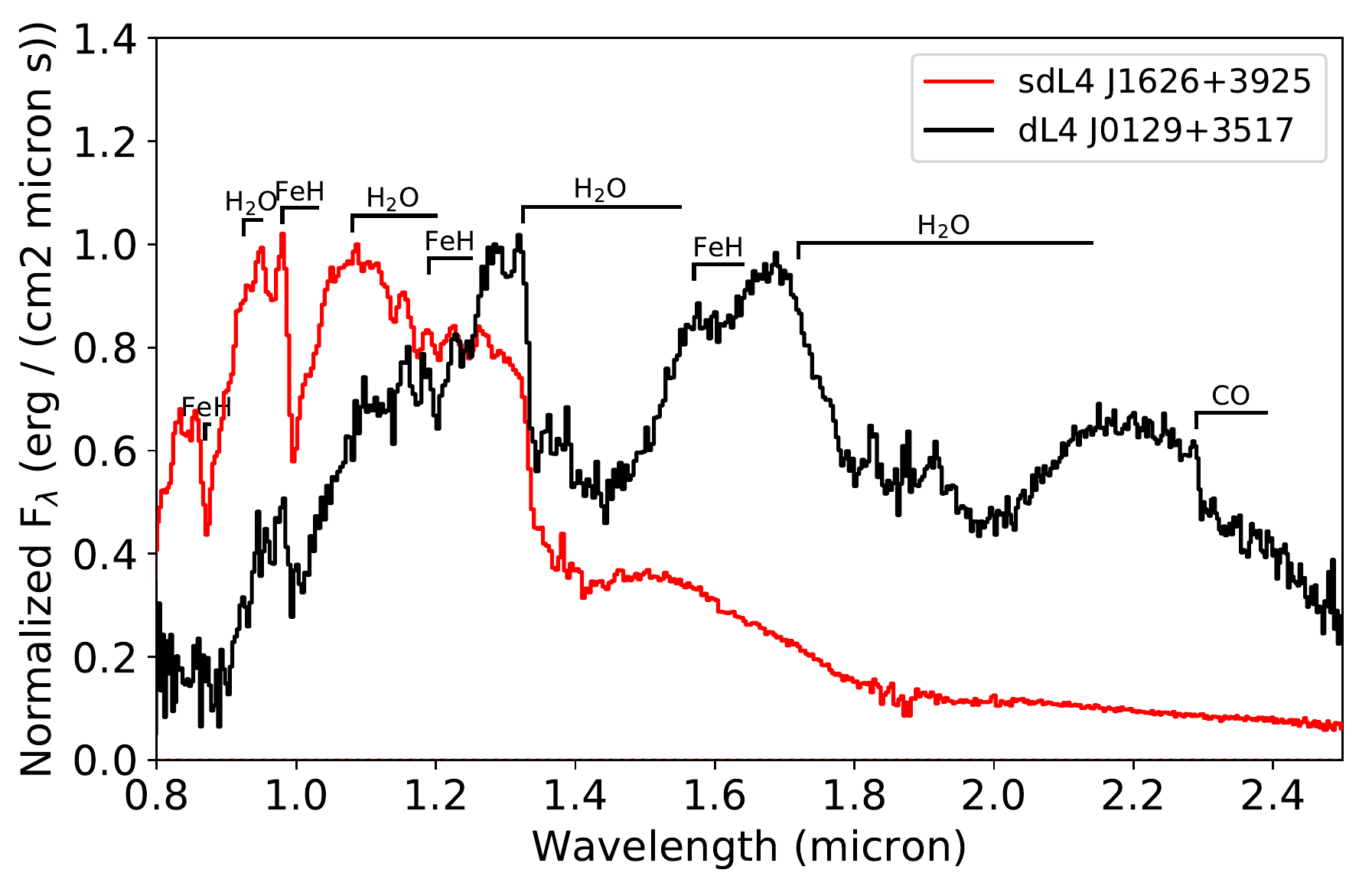}
    \vspace{-0.4cm}
    \caption{\textbf{Left}. Observational optical spectra of a field L4 dwarf (black) and an sdL4 halo subdwarf (red). While similar in shape, the subdwarf exhibits stronger metal hybrid and metal oxide bands, the latter likely a chemical consequence of suppressed condensate formation.
    \textbf{Right} Near-infrared spectra of the same two sources, exhibiting fundamentally distinct spectral energy distributions. The relatively red spectra of field L dwarfs are due largely to condensate dust scattering around 1~$\mu$m. In contrast, dust appears to be absent in the atmospheres of L subdwarfs, which instead exhibit blue infrared spectra shaped by strong collision-induced H$_2$ absorption.}
    \label{fig:spectra}
\end{figure}

\begin{figure}[!htb]
\begin{tabular}{cc}
    \begin{minipage}{0.5\textwidth} \includegraphics[width=\textwidth]{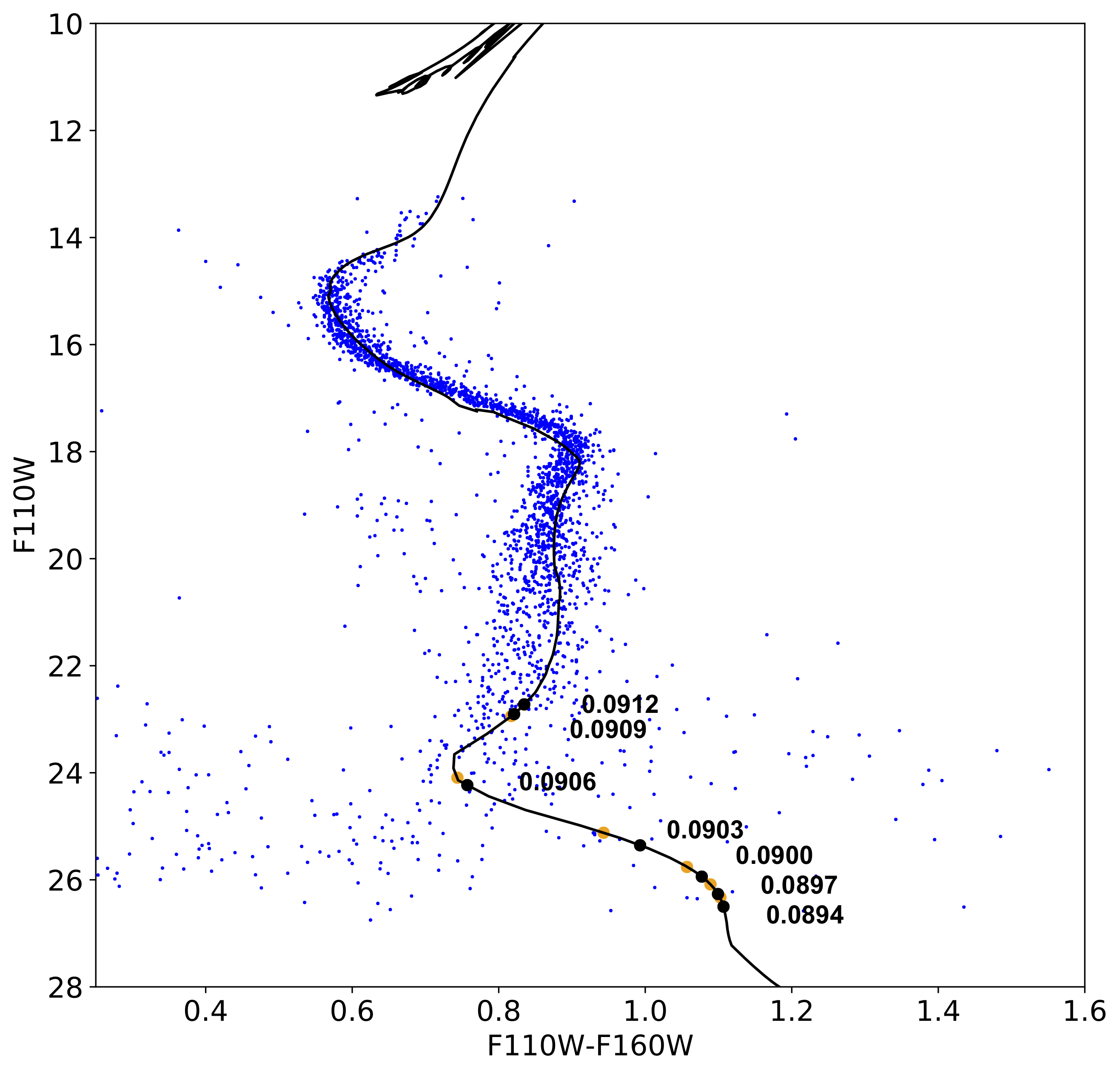} \end{minipage}& \begin{minipage}{0.5\textwidth}  \includegraphics[width=0.9\textwidth]{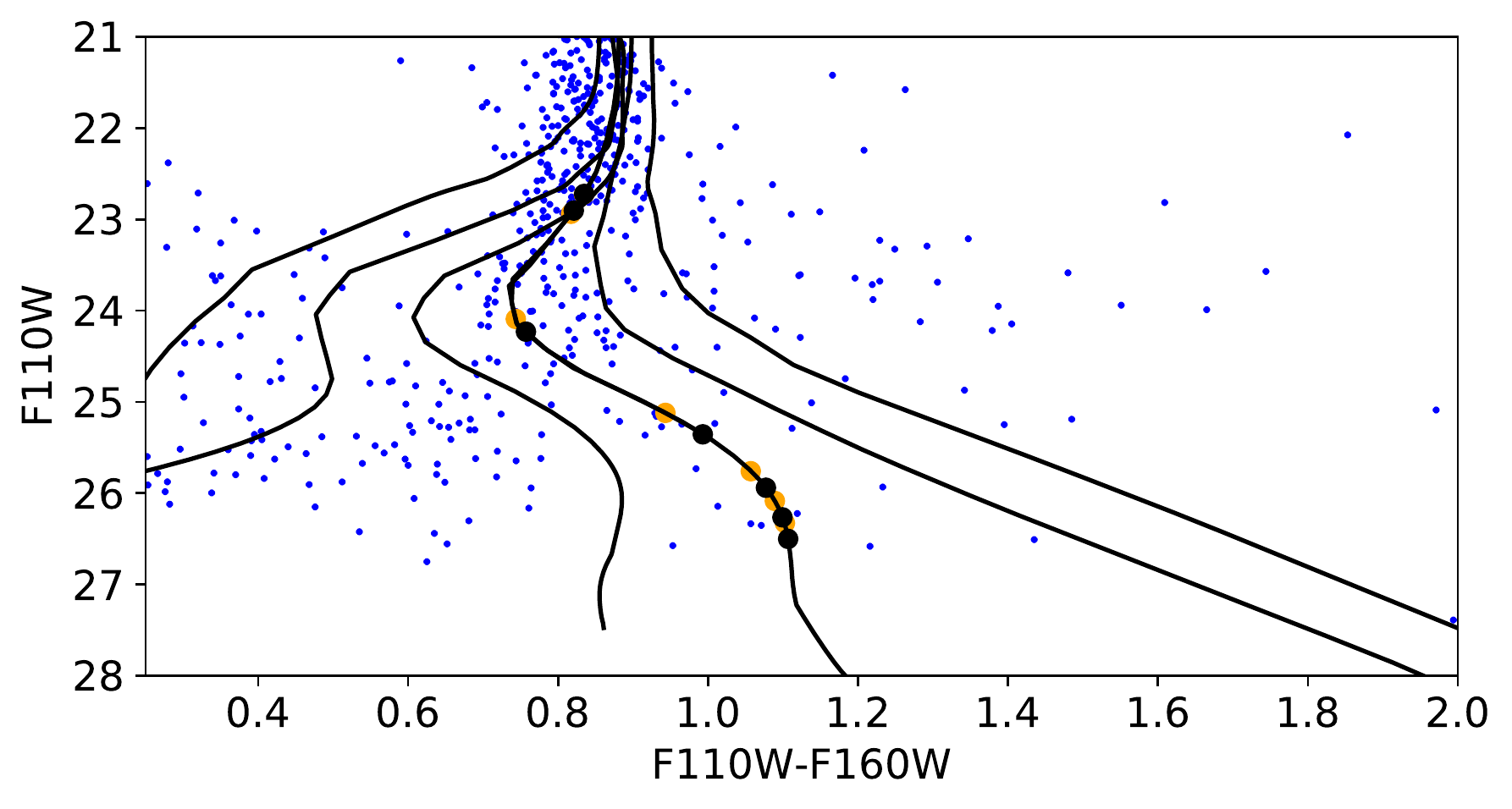} \\
    \includegraphics[width=0.9\textwidth,trim=0.2in 0 0 0]{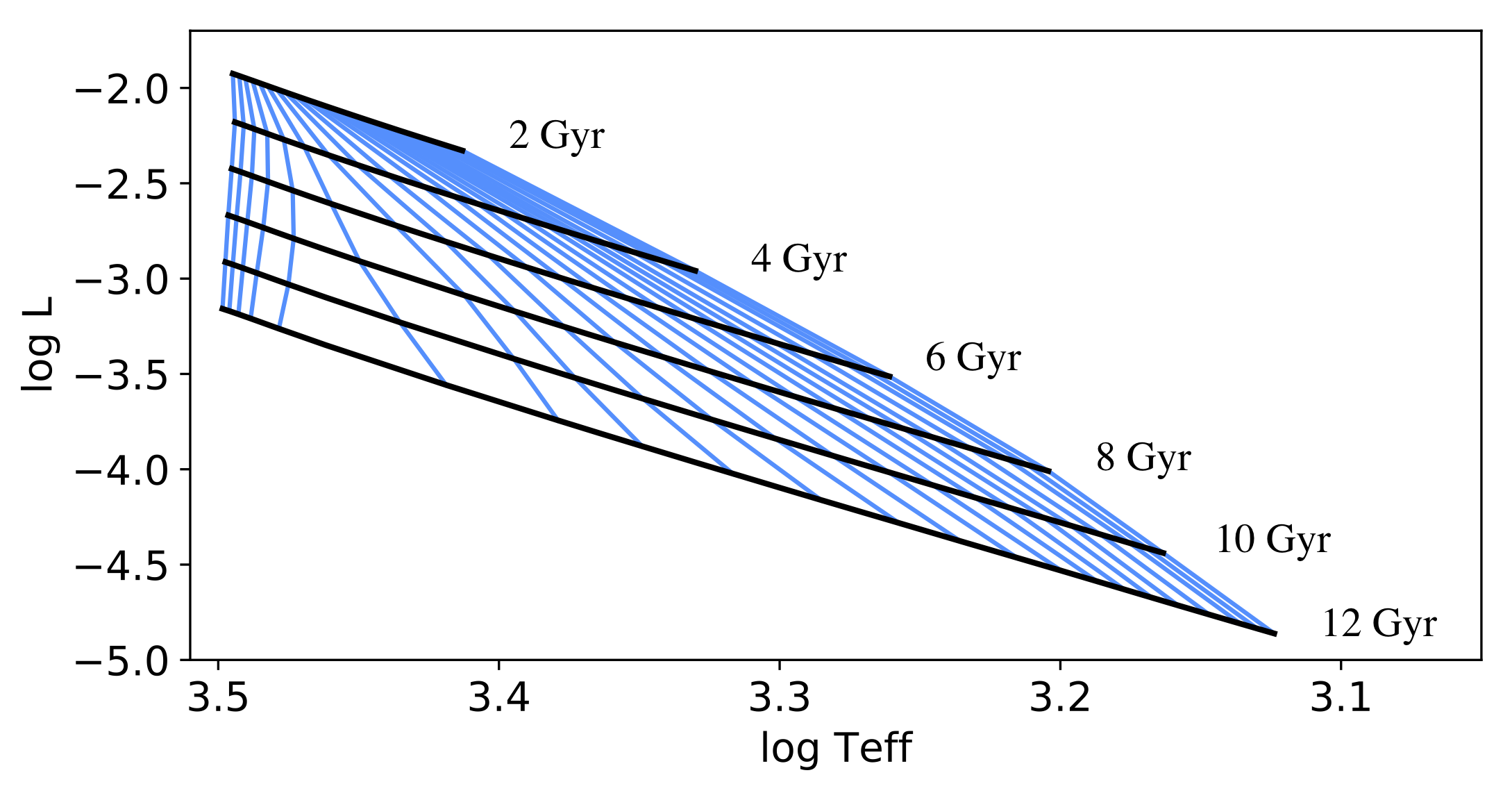} \end{minipage}
    \end{tabular}
        \caption{The left and upper right panels show the gap at the end of the main sequence on the CMD for globular cluster M4; the lower right panel shows the gap opening up with time in the HR diagram. \textbf{Left panel}. Blue dots: data from \cite{2016ApJ...817...48D}; solid black line: 13 Gyr isochrone for [Fe/H]= -1.0, black dots: position at 13 Gyr for 7 masses from 0.0912 to 0.0894 M$_\odot$ in steps of 0.0003 M$_\odot$; orange dots: position for the same masses but at 12 Gyr. We can see from the displacement of the black dots from the orange dots that the low mass stars at the top of the gap (0.0912 and 0.0909 M$_\odot$) get slightly brighter in 1 Gyr, while objects in the gap and brown dwarfs at the bottom get fainter. \textbf{Upper right panel}. Zoom on the end of the main sequence and the gap.  The additional black solid lines represent isochrones at 13 Gyr combined with atmosphere models at different metallicities, from left to right: [Fe/H] = -2.5, -2.0, -1.5, -1.0, -0.5, 0.0. \textbf{Lower right panel}. The solid black lines are isochrones for [Fe/H] = -1.0 at different ages, from 2 Gyr at the top to 12 Gyr at the bottom in steps of 2 Gyr. Except for the 12 Gyr isochrone, all the others have been shifted up, adding 0.25 in $\log L$ every 2 Gyr. The vertical blue lines indicate the paths for 22 masses, from 0.0913 M$_\odot$ (leftmost) to 0.0891 M$_\odot$ (rightmost), evenly spaced in steps of 0.0001 M$_\odot$. The difference in mass between the models is extremely small, and therefore we can see them almost overlapping at the bright end of the gap, where the stars are burning hydrogen and getting slightly hotter with time, and at the faint end, where there is no hydrogen burning and the brown dwarfs have been cooling since the beginning of time. In between, a gap opens where a very small range of masses represent the ``undecided'' objects, with some hydrogen burning in their core but not enough for preventing them from cooling. Evolution models and isochrones calculated with MESA \cite{PaxtonEtal2011,PaxtonEtal2013,PaxtonEtal2015}, atmosphere models by \cite{BaraffeEtal2015,2016sf2a.conf..223A}.}
        \label{fig:CMD}
    \end{figure}

The main difference between a star and a brown dwarf is that hydrogen burning in a brown dwarf will never be the main source of energy or pressure, and cannot stop the brown dwarf from cooling down; while as a star reaches the main sequence, hydrogen burning becomes the main source of energy and the stars increases slowly in temperature and luminosity.  
In a globular star cluster, the lowest-mass main-sequence star and highest-mass brown dwarf approach the main sequence at similar times, but as the former has been getting brighter ever since, the latter has been cooling down and getting fainter. Therefore, at the low mass end of the main sequence we expect to see a gap, populated by very few ``undecided'' objects, which have some hydrogen burning happening in their core, but not enough to keep them from cooling (see Fig.~\ref{fig:CMD}). The mass at which this gap opens is called \textit{hydrogen-burning limit}. This limit is not universal, and presumably depends on composition: a higher metallicity leads to a larger opacity in the atmosphere, causing a higher central temperature and a lower surface luminosity. Current theories predict the hydrogen burning mass limit to range from about 0.07-0.075 solar masses at solar metallicity to about 0.09 M$_\odot$ for very low metal content \cite{1963PThPh..30..460H,2000ARA&A..38..337C,2001RvMP...73..719B}. However, the hydrogen-burning limit has so far not been mapped versus metallicity. This presents one of the major current observational challenges. Globular clusters, spanning a large range of metallicities, provide a great environment to test current theories on how this limit is affected by composition.

\section{Brown dwarfs in globular clusters: learning more about the clusters}

In order to better constrain the unknowns in the physics of stellar evolution and brown dwarfs by comparing our models with observations of globular star clusters, we need to determine with great precision the characteristics --- age, distance from Earth and metallicity --- of the cluster themselves. The age of globular clusters is of particular interest because they are among the oldest objects in our Galaxy, and therefore determining their age can help us understanding the earliest phases of galaxy formation together with setting a lower limit to the age of the Universe \cite{2019arXiv190207081J}. The gap at the low-mass end of the main sequence gets wider with age: from the onset of hydrogen burning, very-low-mass stars get slowly brighter with time, while brown dwarfs never stop cooling down and getting fainter (see Fig.~\ref{fig:CMD}). The width of the gap could therefore provide a new measure of the age of star clusters \cite{2017arXiv170200091C}, independent from commonly used aged indicators like the main sequence turn-off fitting or the detection of the coolest white dwarfs \cite{2005AJ....130..116D,1999AJ....118.2306R,2011ApJ...738...74D,2013Natur.500...51H}. 

From Fig.~\ref{fig:CMD} we can see that the density of stars in the CMD as function of luminosity goes down at the end of the main sequence and picks up again at the lower end of the gap, where the brown dwarf cooling sequence starts. The range in masses over which this occurs is extremely small, of the order of a thousandth of a solar mass (about a two-percent relative change in mass). We can therefore safely extrapolate the mass function that we measure at the lower end of the main sequence to predict how many stars we expect in the gap and at which magnitude we expect the density to increase again depending on the age. This means that we can measure the age of the clusters simply by counting stars as function of magnitude.

The stars' composition in globular clusters is known to a good accuracy, but many clusters present a spread in metallicity and helium content \cite{1999Natur.402...55L,2004ARA&A..42..385G,2005ApJ...621..777P,2007ApJ...661L..53P,2018ARA&A..56...83B}, which in principle could influence the width of the gap. In order to understand how big this effect could be, we consider the case of Omega Centauri ($\omega$ Cen), the cluster with the most striking presence of a spread in composition. Stellar evolution models performed with MESA \cite{PaxtonEtal2011,PaxtonEtal2013,PaxtonEtal2015} for the abundances measured for $\omega$ Cen in \cite{Marino2012}, indicate that 
despite the difference in the hydrogen burning minimum mass due to metallicity, the width of gap between the stellar and substellar populations is relatively insensitive to composition effects. At an age of 5 Gyr, the gap varies between compositions with solar helium abundance ($Y=0.24$) and with supersolar helium abundance of $Y=0.40$ by only 0.06 dex in luminosity, and 50~K in $T_{\mathrm{eff}}$. Even at 13.5 Gyr, the age of $\omega$~Cen estimated by \citet{Milone2017}, the difference between the two compositions is only 0.2~dex in luminosity and 100~K in $T_{\mathrm{eff}}$.  Given that $\omega$ Cen is unusual in having such a large spread in composition, we should expect the effects in other globular clusters to be even smaller.  On the other hand, the luminosity and temperature of stars at the hydrogen-burning limit increases as the metallicity decreases, so finding the upper end of the gap will be easier in the globular clusters and different populations within a particular cluster will exhibit gaps that start at different magnitudes.
We can see from the upper left panel of Fig.~\ref{fig:CMD} (and Fig.~\ref{fig:spectra}) that, due to atmospheric effects, the infrared color of brown dwarfs is extremely sensitive to metallicity in the infrared. This will have to be taken into account when 
determining the age of the cluster, and it can be a tool to study and understand the origin of multiple populations in star clusters.

\section{Where to look}

Globular star clusters tend to dynamically sort stars by their mass. 
Stellar interactions over billions of years bring the more massive stars toward the center and the less massive towards the outskirts. This phenomenon, called mass segregation, is an important factor in determining at which radius within a cluster to observe when looking for brown dwarfs. The center of a cluster is expected to be depleted of low-mass objects, and to be extremely crowded in optical to infrared wavebands, making the detection of the faintest objects nearly impossible. Outer fields, with less crowding and longer relaxation times are therefore preferable, with the caveat that too far from the center the lowest-mass objects have likely evaporated or been ripped away by the Galactic tidal field. Notwithstanding these limitations many clusters are suitable for observing brown dwarfs. For example, deep observations in an outer field of 47 Tuc show an abundance of objects at the end of the main sequence \cite{1538-3881-143-1-11,2013ApJ...778..104R} and by extrapolating the mass function by less than one percent of a solar mass, we expect more than a thousand sub-stellar objects in just two \textit{JWST} fields \cite{2017arXiv170200091C}.

\pagebreak
{\noindent \textbf{Acknowledgments}}\\
We used the NASA ADS service and arXiv.org. We would like to thank Shrinivas Kulkarni for his useful comments.

\bibliographystyle{jer}
\bibliography{main}

\end{document}